\documentclass[pra,aps,showpacs,twocolumn]{revtex4}
\date{\today}
\usepackage{graphicx}

\begin{document}

\title{Restoring integrability in one-dimensional quantum gases by two-particle correlations}
\author{I. E. Mazets$^{1,2}$  and J. Schmiedmayer$^1$}
\affiliation{$^1$Atominstitut der \"Osterreichischen Universit\"aten,
Stadionallee 2, 1020 Vienna, Austria \\
$^2$Ioffe Physico-Technical Institute, 194021 St.Petersburg, Russia}

\begin{abstract}

We show that thermalization and the breakdown of integrability in
the one dimensional Lieb-Liniger model caused by local three-body
elastic interactions is suppressed by pairwise quantum correlations when
approaching the strongly correlated regime.  If the relative
momentum ${k}$ is small compared to the two-body coupling constant
$c$ the three-particle scattering state is suppressed by a factor
of $({k}/c)^{12}$. This demonstrates that in one dimensional
quantum systems it is not the freeze-out of two body collisions
but the strong quantum correlations which ensures integrability.

\end{abstract}

\maketitle

Integrable systems \cite{thack} allow for deep conceptual insight into
many problems of field theory and statistical physics, by
providing exactly solvable models with rich properties.
Thermalization is precluded in integrable models because the number of integrals of
motion  equals to the number of degrees of freedom and
the \emph{'memory'} of the initial state persists in the system
forever. A strictly one-dimensional (1D) system with local (delta-functional) 
interactions is a prime example of an integrable model \cite{ll1}.

In real world, 1D systems are realized by strong
transverse confinement ($\hbar \omega_\perp $ being the level
spacing for the radial confinement Hamiltonian), which
\emph{'freezes out'} the radial degrees of freedom. An example
are ultracold atoms in a tight waveguide where the interval
between its ground and first radially excited states
exceeds both the chemical potential ($\mu \ll \hbar \omega_\perp$)
and the temperature of the atomic system ($k_B T \ll \hbar
\omega_\perp$).  Such a system can be described by the Lieb-Liniger model
\cite{ll1}. Recently it has been shown in experiment that one can cool
in 1D by evaporation to $k_B T < \frac 15{\hbar \omega_\perp}$
\cite{hoff08} and theoretically that the freeze-out of two body
collisions in proportional to $ \exp [-{2 \hbar
\omega_\perp}/({k_B T})]$ does not necessarily lead to a stop of
thermalization \cite{we1}. Local three-body elastic collisions,
where higher radial modes are excited virtually, are only
suppressed by a ratio of the mean interaction energy to $\hbar \omega_\perp $ 
and lead to thermalization and a breakdown of integrability \cite{yur1}.

Nevertheless the beautiful experiments by Kinoshita \emph{et al.}
\cite{weiss} showed that for strongly correlated 1D systems
thermalization is negligible. In this Letter we analyze how these
two experiments and the theoretical model can be reconciled and
show that the virtual three-body collisions described in
\cite{we1} are suppressed by quantum correlations in strongly
correlated 1D Bose gas.

We start by considering $N$ identical bosons in 
1D configuration with the Hamiltonian
\begin{eqnarray}
\hat{H}&=&-\sum _{j=1}^N \frac {\partial ^2}{\partial x_j^2} + 2c
\sum _{j>j^\prime } \delta (x_j-x_{j^\prime }) + \nonumber   \\ &&
\sum _{j>j^\prime >j^{\prime \prime }}U_\mathrm{3b}
(x_j-x_{j^\prime },\, x_j-x_{j^{\prime \prime }}), \label{eq1}
\end{eqnarray}
thereby we use units where Planck's constant is 1 and the mass of
the atom is $\frac 12$. In these units the strength of interaction
of two atoms in a tight waveguide is $c=2a_s/l_\perp ^2$, where
$a_s$ is the three-dimensional (3D) $s$-wave scattering length and
$l_\perp \gg a_s$ is the size of the transverse motion ground
state of an atom in the waveguide \cite{olsh}.

Eq. (\ref{eq1}) differs from the Hamiltonian of the Lieb-Liniger
model by including the term $U_\mathrm{3b}$ that describes
three-body collisions (its value is non-zero, only if both its
arguments are approximately 0, i.e., three atoms come close to
each other). $U_\mathrm{3b}$ is obtained by adiabatic elimination
of transverse modes virtually excited by the 3D short-range
pairwise atomic interaction \cite{we1}.    Note, that effective three-body
interactions between polar molecules emerge in a similar way, due
to virtual transitions to an off-resonant internal state
\cite{z3b}. 

In a non-degenerate state, where free phase space is
available for the scattered particles, $U_\mathrm{3b}$ leads to thermalization
due to scattering of particles, since a three-particle elastic
collision in 1D, unlike a two-particle collision, allows for 
redistribution of the colliding particles momenta. In the ground
state, this type of interactions not only changes the energy of
the system, but also precludes the Lieb-Liniger quasi-momenta from
being integrals of motion. $U_\mathrm{3b}$
is the source of thermalization and leads to a
beakdown of integrability.

In what follows, we calculate this three-body scattering amplitude
in the presence of the delta-functional pairwise interactions. The
real part of this scattering amplitude determines the vertex for
the effective three-body interaction by
analogy with two-particle scattering in two dimensions
\cite{schick, shl2da, maz11}. The thermalization rate is proportional
to the square of its absolute value. The stronger the
{\em pairwise} interparticle
repulsion, the smaller is the probability of a close encounter of three
particles. This results in suppression of the three-body
scattering amplitude, which is the main subject of the present Letter.

To analyze the Hamiltonian (\ref{eq1}) for $N=3$ we express it
in hyperspherical coordinates $R$, $\alpha $ defined as \cite{o1}
\begin{equation}
R \sin \alpha =\frac {x_1-x_2}{\sqrt{2}} , \quad
R \cos \alpha =\sqrt{\frac 23}\left( x_3-\frac {x_1+x_2}2\right)
\label{eq2}
\end{equation}
and the center-of-mass coordinate $X=\frac 13(x_1+x_2+x_3)$:
\begin{eqnarray}
\hat{H}&=&-\frac 13\frac {\partial ^2}{\partial X^2} -\frac 1R
\frac \partial {\partial R}R\frac \partial {\partial R}
-\frac 1{R^2}\frac {\partial ^2}{\partial \alpha ^2} +\nonumber \\ &&
\frac {\sqrt{2}c}R\sum _{\nu =-2}^3 \delta (\alpha -\nu \pi /3) +
U_\mathrm{3b} (R,\alpha ).
\label{eq3}
\end{eqnarray}
If one deals with three particles in the 3D space, the total
number of degrees of freedom equals to 9. After separation of the
center-of-mass motion, six degrees of freedom remain. The
corresponding coordinates in the usual 3D case are $R$, $\alpha $
and four angles characterizing the orientation of two independent
differences of the particles radius-vectors \cite{o1}. In our 1D
case, the co-ordinate space reduces, after separation of the
center-of mass-motion, to a plane, $R$ and $\alpha $ being its
polar co-ordinates.

\begin{figure}[t]
\includegraphics[width=6.0cm]{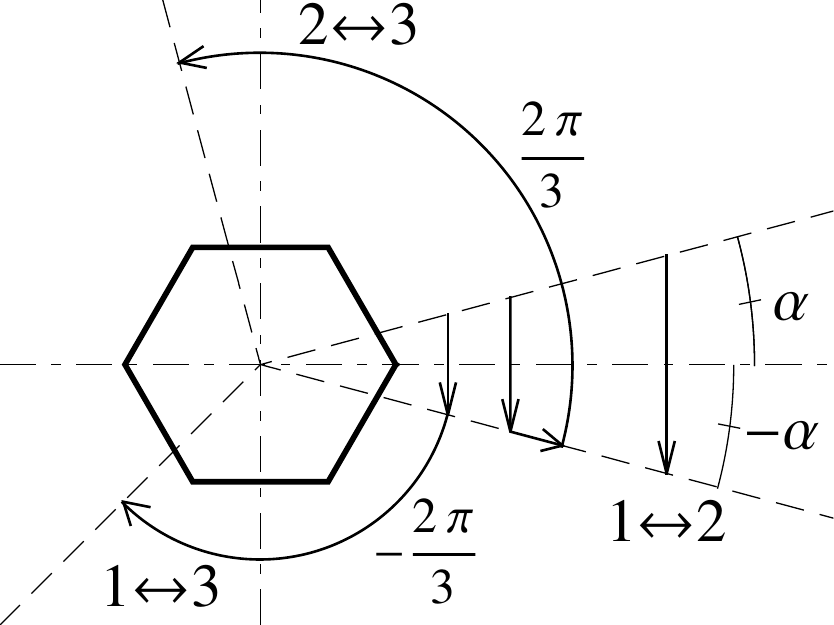}
\vspace*{3mm}
\begin{caption}
{Transformations of the hyperangle $\alpha $ corresponding to the
pairwise permutations of particles $j\leftrightarrow j^\prime $.
The regular hexagon indicates the symmetry of the Hamiltonian of
three pairwise-interacting identical particles. The two-body
interaction potential has delta-functional singularity on straight
lines (not shown in this figure) going through opposite vertices
of the hexagon, cf. Eq. (\ref{eq3}). See text for the  details.}
\end{caption}
\end{figure} 

For indistinguishable bosons the permutation
$x_1\leftrightarrow x_2$ corresponds to the reflection operation
that changes $\alpha $ to $-\alpha $. The two remaining pairwise
permutations, $x_3\leftrightarrow x_2$ and $x_1\leftrightarrow
x_3$, correspond to the product of this reflection and rotation in
the plane by $-2\pi /3$ and $2\pi /3$, respectively, as shown in
Fig.~1. The two cyclic permutations of $x_1$, $x_2$, $x_3$
correspond to mere rotations by $\pm 2\pi /3$. The three-body
interaction potential (its particular form will be specified
later) must be  invariant with respect to the permutation of the
particles. The Hamiltonian of the three particles interacting via
pairwise interactions only, i.e., Eq. (\ref{eq3}) except of the
term $U_\mathrm{3b} (R,\alpha )$, possesses a richer symmetry
(the symmetry group ${\bf D} _6$ of a regular hexagon \cite{d6};
two of its six reflection-symmetry axes  are shown in Fig.~1 by
dash-dotted lines). Adding any reasonable effective three-body 
interaction does not change this symmery.

The Schr\"odinger equation for the three-particle wave function is
$\hat{H}\Psi (x_1,x_2,x_3) =(k_1^2+k_2^2+k_3^2)\Psi
(x_1,x_2,x_3)$. The wavenumbers $k_j$ can be defined from the set
of transcendental equations \cite{ll1}, provided that the periodic
boundary conditions are set on the interval of the length $L$. By
setting $L\rightarrow \infty $, we obtain a continuous spectrum of
real $k_j$'s in the repulsive case ($c>0$). Then we separate the
center-of-mass motion and describe the relative motion in
hyperspheric coordinates: $\Psi (x_1,x_2,x_3)=\exp [
i(k_1+k_2+k_3) X] \psi _\mathrm{r}(R,\alpha )$. The kinetic energy
of the relative motion is $k^2=\frac 13 [
(k_1-k_2)^2+(k_2-k_3)^2+(k_3-k_1)^2]$. We decompose the wave
function of the three colliding particles in the center-of-mass
frame as  \cite{o1}
\begin{equation}
\psi _\mathrm{r}(R,\alpha )=\sum _{n=0}^\infty F_n(R)\chi _n(R,\alpha ),
\label{eq6}
\end{equation}
where $\chi _n(R,\alpha )$ is an eigenfunction of the hyperangle-dependent
part of the Hamiltonian:
\begin{eqnarray}
&&\left[-\frac {\partial ^2}{\partial \alpha ^2}+\sqrt{2}cR
\sum _{\nu =-2}^3 \delta (\alpha -\nu \pi /3)\right] \chi _n(R,\alpha )=
\qquad \nonumber \\ && \qquad \qquad \qquad \qquad \qquad \qquad
\qquad \lambda _n^2(R)\chi _n(R,\alpha ).
\label{eq7}
\end{eqnarray}
Due to the bosonic symmetry,
\begin{equation}
\chi _n(R,\alpha )=\chi _n(R,\pm \alpha + 2\pi \nu ^\prime /3),
\qquad \nu ^\prime =-1,0,1.
\label{eq8}
\end{equation}
Eq. (\ref{eq8}) defines the spectrum of eigenvalues and, hence, the
adiabatic hyperspheric potentials $\lambda _n^2(R)/R^2$. The eigenfunctions are
\begin{equation}
\chi _n(R,\alpha )=\tilde{\chi }_n(R,\alpha )+\tilde{\chi }_n(R,\alpha -2\pi /3)+
\tilde{\chi }_n(R,\alpha +2\pi /3),
\label{eq9}
\end{equation}
where
\begin{equation}
\tilde{\chi }_n(R,\alpha )=\left \{ \begin{array}{ll}
\cos [\lambda _n(R)(\pi /6-|\alpha |)-\pi n/2] , & | \alpha |\leq \pi /3 \\
0 ,  & \mathrm{otherwise} \end{array} \right.
\label{eq10}
\end{equation}
and the eigenvalues are the positive roots of the equation
\begin{equation}
\lambda _n(R) \tan [ \pi \lambda _n(R)/6-\pi n/2]=cR/\sqrt{2} ,~n=0,1,2, \, \dots .
\label{eq11}
\end{equation}

After integrating out the hyperangular variable, the Schr\"odinger equation
reduces to the set of coupled ordinary differential equations
\begin{eqnarray}
&&-\frac 1R \frac d{dR}R\frac d{dR} F_n-\sum _{n^\prime =0}^\infty \left \{
\tilde{W}_{nn^\prime }(R) \frac d{dR} F_n +\right.  \nonumber \\ &&
\left. [\tilde{Y}_{nn^\prime }(R) -\tilde{U}_{nn^\prime }(R)]
F_{n^\prime } \right \}+\frac {\lambda _n^2(R)}{R^2}F_n=k^2F_n ,
\label{eq12}
\end{eqnarray}
where
\begin{eqnarray}
\tilde{W}_{nn^\prime }(R)&=& 2\int _0^{\pi /3}d\alpha \, \frac {\tilde{\chi }_n
(R,\alpha )}{B_n(R)}\frac \partial {\partial R}
\tilde{\chi }_{n^\prime }(R,\alpha ) ,
\label{eq13} \\
\tilde{Y}_{nn^\prime }(R)&=&\int _0^{\pi /3}d\alpha \,
\frac{\tilde{\chi }_{n}(R,\alpha )}{B_n(R)R} \frac \partial {\partial R}R
\frac \partial {\partial R}\tilde{\chi }_{n^\prime }(R,\alpha ) ,
\label{eq14} \\
\tilde{U}_{nn^\prime }(R)&=&\int _0^{\pi /3}d\alpha \,
\frac {\tilde{\chi }_{n}(R,\alpha )U_\mathrm{3b}(R,\alpha )
\tilde{\chi }_{n^\prime }(R,\alpha )}{B_n(R)} , ~~~
\label{eq15} \\
B_n(R)&=&\int _0^{\pi /3}d\alpha \, \tilde{\chi }_{n}^2(R,\alpha ) .
\label{eq16}
\end{eqnarray}
Neglecting all off-diagonal (coupling) terms is termed adiabatic
hyperspherical approximation \cite{mc1}. However, for the sake of
clarity, we neglect also diagonal terms $\tilde{W}_{nn}$
and $\tilde{Y}_{nn}$ \cite{o1}. Indeed,
the respective terms decrease more rapidly than others by a factor of $1/R$ as
$R\rightarrow \infty $, and at $cR\lesssim 1$ the correction to the adiabatic
potential $\lambda _0^2(R)/R^2$ due to $\tilde{Y}_{00}(R)$ is less than 5~\%{.}
The finite value of $\tilde{W}_{00}$ can be neglected compared to
$\lambda _0^2(R)/R^2\rightarrow \infty $ as $R\rightarrow 0$.

In what follows we assume the long-wavelength limit
$k\ll c$, which holds in the regime of strong interactions \cite{ll1}
at temperatures of the order of or less than the chemical potential. It enables us to
consider in Eqs. (\ref{eq6},~\ref{eq12}) only the lowest hyperangular mode, $n=0$,
because in higher modes scattering is not efficient. We introduce the effective
radius $r_0$ of the three-body interactions: $U_\mathrm{3b}$ rapidly vanishes
for interatomic separation values exceeding $r_0$. For atoms trapped in a
tight waveguide, $r_0$, like the effective range of two-body collisions, is much smaller
than other available length scales, in particular, $r_0\ll c^{-1}$.
We consider now, under the assumptions mentioned above, the equation
\begin{equation}
-\frac 1R \frac d{dR}R\frac d{dR} F_0+\left[ \frac {\lambda _0^2(R)}{R^2}+
\tilde{U}_{00}(R)\right] F_0=k^2F_0
\label{eq17}
\end{equation}
with the boundary conditions requiring $F_0$ to be finite for both $R=0$ and
$R\rightarrow \infty $, 
for the ``partial wave" corresponding to the lowest eigenvalue $\lambda _0(R)$, whose
asymptotic expressions are
\begin{equation}
\lambda _0(R)\approx \left \{ \begin{array}{ll}
\sqrt{ \frac {3\sqrt{2}cR}\pi },&cR\ll 1\\
3-\frac {18\sqrt{2}}{\pi cR}, &cR\gg 1\end{array}\right.  .
\label{eq18}
\end{equation}
We assume that the averaged over the hyperangle three-body interaction potential
$\tilde{U} _{00}(R)$ vanishes at $R>r_0$, where $r_0\ll c^{-1}$.

First we solve Eq. (\ref{eq17}) for $R>r_0$. For $cR\lesssim 1$ the two
independent solutions are the modified Bessel functions of the zeroth order \cite{aste}
$I_0(2 \sqrt{\beta R})$ and $K_0(2 \sqrt{\beta R})$, where $\beta =(3
\sqrt{2}/\pi )c$ (we call them the inner solutions). The outer {\em real} solutions
hold for $cR\gg 1$ and are the Bessel functions of the third order $J_3(kR)$ and
$Y_3(kR)$. Now we note,  that in the range
$c^{-1}\ll R\ll k^{-1}$ the outer solutions can be represented by their asymptotics
$J_3(kR)\approx (kR)^3/48$ and $Y_3(kR)\approx (-16/\pi )/(kR)^3$. The outer
and inner solutions can be tailored by quasiclassical solutions in two dimensions
\begin{equation}
\phi _\mathrm{qc}^\pm (R)= \frac {C^\pm }{\sqrt{\lambda _0(R)}} \exp \left[ \pm
\int _0^R dR^\prime \, \frac {\lambda _0(R^\prime )}{R^\prime }\right] .
\label{eq21}
\end{equation}
Note that two-dimensional quasiclassical solutions differ from one-dimensional ones
\cite{ldl} by a prefactor $1/\sqrt{R}$.
By a proper choice of the constants $C^\pm $ we match at $R\sim c^{-1}$ the quasiclassical
solutions to the inner solutions: $\phi _\mathrm{qc}^+ (R)$
to $I_0(2\sqrt{\beta R})\approx (4\pi
\sqrt{\beta R})^{-1/2}\exp (2\sqrt{\beta R})$ and $\phi _\mathrm{qc}^-(R)$
to $K_0(2\sqrt{\beta R})
\approx (4\sqrt{\beta R}/\pi )^{-1/2}\exp (-2\sqrt{\beta R})$, respectively. On the
other hand, the asymptotics of the quasiclassical solutions at $cR\gg 1$ are
\begin{equation}
\phi _\mathrm{qc}^+(R)\approx \frac \Xi {\sqrt{12\pi }}(cR)^3, \quad
\phi _\mathrm{qc}^-(R)\approx \frac {\sqrt{\pi }}{\sqrt{12}\Xi }(cR)^{-3},
\label{eq22}
\end{equation}
where $\Xi \approx 0.18$ is a numerical constant. This tailoring enables us to
construct two independent real solutions $F^\pm (R)$
of Eq. (\ref{eq17}) in the whole range $R>r_0$ with the following asymptotics:
\begin{equation}
F^+(R)\approx \left \{ \begin{array}{ll}
I_0(2\sqrt{\beta R}), & cR\lesssim 1 \\
\frac {8\sqrt{3}\Xi }{\sqrt{\pi }} \left( \frac ck\right) ^3J_3(kR),& cR\gg 1
\end{array} \right. ,
\label{eq23}
\end{equation}
\begin{equation}
F^-(R)\approx \left \{ \begin{array}{ll}
K_0(2\sqrt{\beta R}), & cR\lesssim 1 \\
-\frac {\pi \sqrt{\pi }}{32\sqrt{3}\Xi } \left( \frac kc\right) ^3
[Y_3(kR)+bJ_3(kR)],& cR\gg 1
\end{array} \right. ,
\label{eq24}
\end{equation}
where $b$ is a real number of the order of or less than 1. In other words, the
term $Y_3(kR)\propto R^{-3}$ is only the leading term of the
expansion of $F^-(R)$ in the intermediate range of $R$; the solution
$J_3(kR)$ vanishing for small $kR\ll 1$ is contained, in a general case,
in terms beyond the accuracy of the quasiclassical approximation \cite{ldl}.
The upper limit to the magnitude of $b$
follows from the fact that the quasiclassical expression (\ref{eq22}) for
$\phi _\mathrm{qc}^-(R)$ holds up to $R\sim k^{-1}$.

The general solution of Eq. (\ref{eq17}) for $R>r_0$ is then
\begin{equation}
F_0(R)=C_1F^+(R)+C_2F^-(R).
\label{eq25}
\end{equation}
The particular ratio between the coefficients $C_1$ and $C_2$ is found from
matching the logarithmic derivatives of the solution given by Eq. (\ref{eq25})
and its solution at $R<r_0$.

\emph{As a concrete example}, illustrating the basic physics through 
simple analytic expressions, we take the three-body interaction
potential in the form of the sum of a rectangular potential well and
a term compensating the adiabatic hyperspherical potential at
$R<r_0$:
\begin{equation}
\tilde{U}_{00}(R)=-q^2-\beta /R, \qquad R<r_0.
\label{eq26}
\end{equation}
Without loss of generality, we assume further that $k\ll q$. In this case
the regular solution at $R<r_0$ is $J_0(qR)$. Matching the logarithmic derivatives
of the solutions at $R>r_0$ and $R<r_0$, recalling that $r_0\ll c^{-1}$, and
using the asymptotics $I_0(z)\approx 1$ and $K_0(z)\approx -\ln z$ at
$z\rightarrow 0$, we obtain
\begin{equation}
\frac {C_1}{C_2}=\ln (2\sqrt{\beta r_0})+\frac {J_0(qr_0)}{2qr_0J_1(qr_0)} .
\label{eq27}
\end{equation}
Then we calculate the partial scattering amplitude $\tilde{f}_0$ 
(corresponding to the scattering channel with
$n=0$) introduced by expressing Eq. (\ref{eq25})
at $R\rightarrow \infty $ in the form
\begin{equation}
F_0(R)\approx  \mathrm{const} \, [J_3(kR)-i\tilde{f}_0H^{(1)}_3(kR)] ,
\label{eq28}
\end{equation}
where $H^{(1)}_3(z)=J_3(z)+iY_3(z)$ is the Hankel function of the
first kind, corresponding to the outgoing (scattered) wave. Note
that there are various definition of the scattering amplitude in
2D, differing by a complex prefactor \cite{shl2da,ajp81,ajp86}.
The rate of three-body
collisions is proportional to $|\tilde{f}_0|^2$. 

Using Eqs. (\ref{eq23}, \ref{eq24}), we obtain
\begin{equation}
\tilde{f}_0=\frac 1 { -\Omega \left( \frac ck\right) ^6 \left[ \ln (4\beta r_0)+
{\cal J} \right] +b+i } ,
\label{eq29}
\end{equation}
where $\Omega =384 (\Xi /\pi )^2\approx 1.26$ and ${\cal
J}=J_0(qr_0)/[qr_0J_1(qr_0)]$. Other choice of $\tilde{U}_{00}(R)$
rather than Eq. (\ref{eq26}) gives different value for ${\cal J}$,
but does not change the general structure of Eq. (\ref{eq29}).
Although the exact value of $b$ can not be obtained from the
quasiclassical matching of solutions, it is relatively small ($|b|\lesssim 1$) and
thus $b$ can be neglected in Eq. (\ref{eq29}).

From  Eq. (\ref{eq29}) we conclude that the three-body scattering
amplitude decreases in proportion to $(k/c)^6$ as $c\rightarrow
\infty $, and the three-body scattering rate in a 1D system of
bosons in the case of strong pairwise interaction is suppressed by
a factor $\sim (k/c)^{12}$. This suppression is the main result of
the present paper. Averaging over collision momenta in a
moderately-excited strongly-interacting state we obtain the
scattering rate suppression factor $\sim \gamma ^{-12}$, where
$\gamma =c/n_{1D}$ is the Lieb-Liniger parameter \cite{ll1} and
$n_{1D}=\langle \hat{\Psi }^{\dag }(x)\hat{\Psi }(x)\rangle $ is
the 1D density of particles, $\hat{\Psi }(x)$ being the bosonic
field annihilation operator.

We can now compare our result with the zero-distance
three-particle correlation function $g_3(0)=\langle \hat{\Psi
}^{\dag \, 3}(x)\hat{\Psi }^{ 3}(x)\rangle /n_{1D}^3$.
In the strong interation limit $\gamma \gg 1$,
where $g_3(0)\propto \gamma ^{-6}$  \cite{g3a,g3b}.  This thus proves the
conjecture \cite{we1} that the pairwise interactions and the
quantum correlations induced by them in a strongly-interacting 1D
bosonic system suppress the three-body elastic scattering rate,
and, hence, thermalization, by a factor $\propto g_3^2(0)$. In other
words, strong quantum correlations restore integrability as a 1D system
approaches the Tonks-Girardeau regime. This observation is in
agreement with the experimental results \cite{weiss}.

In conclusion we remark, that integrability in tightly
confined 1D systems is not, as suggested by quasiclassical
arguments, caused by the freeze-out of two-body collisions, but is
ensured by the quantum correlations in a strongly interacting 1D
system. The genuinely quantum effect of virtual excitations to higher radial 
states opens a way to thermalization by three-body collisions and breaks 
integrability in the quantum description of a 1D system.  These three-body 
collisions can be suppressed by quantum correlations caused by strong pairwise 
repulsions. If they dominate, as in a strongly correlated 1D Tonks-Girardeau 
gas, they restore the integrability of the system.
This points to an interesting difference between
classical and quantum systems.

This work is supported by the EC (STREP MIDAS) and the FWF.
I.E.M. acknowledges  support through the Lise Meitner program by the FWF. We thank 
Prof. D. V. Fedorov for helpful discussions.

\end{document}